\documentclass[sigconf,anonymous=false]{acmart}

\usepackage{xspace}
\usepackage{bbm}
\usepackage{amsmath}
\usepackage{svg}
\usepackage{xcolor,cancel}
\usepackage{mathtools}
\usepackage{dsfont}
\usepackage{wrapfig}

\usepackage{caption}
\usepackage{subcaption}
\usepackage{eucal}

\usepackage{array}

\makeatletter
\def\Hline{
  \noalign{\ifnum0=`}\fi\hrule \@height 4.\arrayrulewidth \futurelet
  \reserved@a\@xhline}
\makeatother

\usepackage{amsmath}
\usepackage{graphicx}
\usepackage{here}
\PassOptionsToPackage{hyphens}{url}
\usepackage{hyperref} 
\usepackage{url} 
\usepackage{textcomp} 
\usepackage{multicol}
\usepackage{subcaption}
\usepackage{comment}
\usepackage{booktabs,dcolumn}
\usepackage{multirow}
\usepackage{siunitx} 
\usepackage{fancyhdr}			
\usepackage{totpages}
\usepackage{arydshln}
\usepackage{enumitem}
\usepackage{svg}
\usepackage{rotating}
\usepackage[stable]{footmisc}
\usepackage{soul}
\def\etal{\textit{et al.}}
\def\ie{\textit{i.e.}}

\def\datasetname{\textit{HuTics}\xspace}
\def\systemname{LookHere\xspace}
\def\vimt{V-IMT\xspace}

\renewenvironment{quote}[1][0.04\linewidth]
  {\list{}{\leftmargin=#1\rightmargin=#1}\item\relax}{\endlist}
\newcommand{\myquote}[2]
{
\begin{quote}
\textit{``#1''} [#2]
\end{quote}
}

\usepackage{color}
\definecolor{orange}{RGB}{255,127,0}
\definecolor{darkgreen}{RGB}{0, 146, 0}
\definecolor{violet}{RGB}{148,0,211}

\newcolumntype{L}[1]{>{\raggedright\let\newline\\\arraybackslash\hspace{0pt}}m{#1}}
\newcolumntype{C}[1]{>{\centering\let\newline\\\arraybackslash\hspace{0pt}}m{#1}}
\newcolumntype{R}[1]{>{\raggedleft\let\newline\\\arraybackslash\hspace{0pt}}m{#1}}

\newif\ifCOMMENTS
\COMMENTSfalse
\ifCOMMENTS

\newcommand{\major}[1]{\hl{#1}}

\else

\newcommand{\major}[1]{#1}

\fi

\makeatletter
\def\Hline{
  \noalign{\ifnum0=`}\fi\hrule \@height 4.\arrayrulewidth \futurelet
   \reserved@a\@xhline}
\makeatother

\AtBeginDocument{%
  \providecommand\BibTeX{{%
    \normalfont B\kern-0.5em{\scshape i\kern-0.25em b}\kern-0.8em\TeX}}}

\copyrightyear{2022}
\acmYear{2022}
\setcopyright{acmcopyright}\acmConference[UIST '22]{The 35th Annual ACM Symposium on User Interface Software and Technology}{October 29-November 2, 2022}{Bend, OR, USA}
\acmBooktitle{The 35th Annual ACM Symposium on User Interface Software and Technology (UIST '22), October 29-November 2, 2022, Bend, OR, USA}
\acmPrice{15.00}
\acmDOI{10.1145/3526113.3545648}
\acmISBN{978-1-4503-9320-1/22/10}

\begin{document}

\title[Gesture-aware Interactive Machine Teaching with In-situ Object Annotations]{Gesture-aware Interactive Machine Teaching\\with In-situ Object Annotations}




\author{Zhongyi Zhou, Koji Yatani}
\affiliation{
  \institution{Interactive Intelligent Systems Lab., The University of Tokyo}
  \city{Tokyo}
  \country{Japan}
}
\email{{zhongyi, koji}@iis-lab.org}

\begin{abstract}
Interactive Machine Teaching (IMT) systems allow non-experts to easily create Machine Learning (ML) models.
However, existing vision-based IMT systems either ignore annotations on the objects of interest or require users to annotate in a post-hoc manner.
Without the annotations on objects, the model may misinterpret the objects using unrelated features.
Post-hoc annotations cause additional workload, which diminishes the usability of the overall model building process.
In this paper, we develop LookHere, which integrates in-situ object annotations into vision-based IMT.
LookHere exploits users' deictic gestures to segment the objects of interest in real time.
This segmentation information can be additionally used for training.
To achieve the reliable performance of this object segmentation, we utilize our custom dataset called \textit{HuTics}, including 2040 front-facing images of deictic gestures toward various objects by 170 people.
The quantitative results of our user study showed that participants were 16.3 times faster in creating a model with our system compared to a standard IMT system with a post-hoc annotation process while demonstrating comparable accuracies.
Additionally, models created by our system showed a significant accuracy improvement ($\Delta mIoU=0.466$) in segmenting the objects of interest compared to those without annotations.


\end{abstract}



\begin{teaserfigure}
  \includegraphics[width=\textwidth]{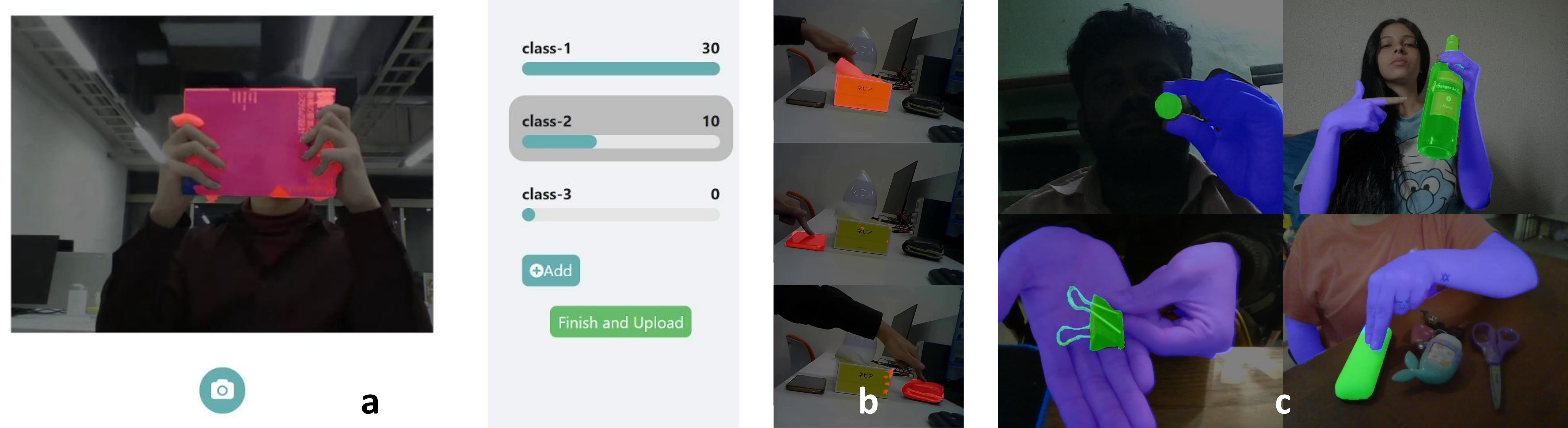}
  \caption{
  (a): The teaching interface of our vision-based Interactive Machine Teaching system, \systemname. \systemname provides a segmentation mask (an object highlight) on the object guided by users' deictic gestures in real time during teaching. This segmentation mask is used for model training as additional information for training classifiers.
  (b): Users' deictic gestures guides in-situ object annotations.
  (c): Example images in our \datasetname dataset that enables the implementation of LookHere. \datasetname~includes 2040 labeled images that capture how 170 people use deictic gestures to present an object.
  }
  \Description{xx}
  \label{fig:teaser}
\end{teaserfigure}

\settopmatter{printfolios=true}

\maketitle


\section{Introduction}

Interactive Machine Teaching (IMT)~\cite{ramos2020imt, simard2017machine} aims to enhance users' teaching experience during the creation of Machine Learning (ML) models.
IMT systems are primarily designed for non-ML-experts, and allow such users to provide training data through demonstrations.
Vision-based IMT (\vimt) systems utilize cameras to capture users' demonstrations.
For example, in Teachable Machine~\cite{Carney2020teachable} users can create a computer vision classification model by showing different views of each object (class) to a camera.
Despite its low burden for providing training samples, existing work~\cite{zhou2021enhancing} revealed that an ML model trained through a \vimt system might recognize an object by using visual features unrelated to it.
For example, even if a user performs a demonstration of a book, the model may use visual features in the background.
Failure to address this error properly could result in the degraded performance of the model when it is deployed to real applications.
Therefore, users should have the capability to specify portions of an image that a model should emphasize in learning to achieve reliable classification. 


One approach to address this issue is to perform annotations on the objects of interest~\cite{boris2020cvat} and feed them into the model as well.
Advances in annotation tools reduce user workload by simplifying necessary interactions to clicks~\cite{mortensen1998interactive, fbrs2020sofiiuk} or sketches~\cite{rother2004grabcut, Zhang2020weakly}. 
Despite their lowered burden, existing annotation tools are not well tailored toward \vimt systems, and users thus have to perform annotations in a post-hoc manner.
This would degrade the overall experience of \vimt systems~\cite{smith2020no}.
Annotation approaches that are more deeply integrated into \vimt thus need to be explored.

To this end, this work examines \vimt systems that can integrate object annotations into the teaching process.
We observe that when users are doing demonstrations for teaching, they may hold or point to the object of interest.
These deictic gestures in demonstrations thus are indicative of what visual features a model should focus on.
Therefore, this work focuses on the integration of annotations by leveraging deictic gestures that humans naturally perform during the teaching process.
Note that, in this paper, we use the term deictic gestures to represent a wide range of gestures whose purpose is to indicate the object of interest~\cite{sauppe2014robdeic, clark2005coordinating} while it typically represents pointing gestures in HCI research~\cite{karam2005taxonomy}.

Our \vimt system, called \systemname\footnote{\major{The source code is available at} \url{https://github.com/zhongyi-zhou/GestureIMT}}, embeds in-situ object annotations inferred from users' deictic gestures into the teaching process.
\systemname provides real-time visualizations, named \textit{object highlights}, on what portions of the given frame the system is considering as the region of the target object (the red mask in Figure~\ref{fig:teaser}a).
Depending on the deictic gestures users are performing, our system infers different object regions (Figure~\ref{fig:teaser}b). 
To achieve this gesture-aware object segmentation, we created \datasetname, a dataset consisting of 2040 images collected from 170 people that include various deictic gestures and objects with segmentation mask annotations (Figure~\ref{fig:teaser}c).
Our technical evaluation shows that our object highlights can achieve the accuracy of 0.718 (mean Intersection over Union; $mIoU$) and can run at 28.3 fps.
Our user evaluation confirms that participants were able to build accurate models while being liberated from post-hoc manual object annotations.

This work offers the following contributions:
\setlength{\leftmargini}{10pt}
\begin{itemize}
    \item A vision-based IMT system, \systemname, which integrates in-situ object annotations guided by users' deictic gestures into the teaching process,
    \item The development of real-time object highlights, which offers users feedback on the object region inferred from their deictic gestures,
    \item The \datasetname dataset\footnote{\major{The dataset is available at} \url{https://zhongyi-zhou.github.io/GestureIMT/}.}, which contains 2040 labeled images from 170 participants interacting with various target objects using deictic gestures, and
    \item Our evaluations that confirm \systemname's benefits through quantitative and qualitative results.
\end{itemize}
\section{Related Work}
\subsection{Interactive Machine Teaching}
Machine Teaching is a term that has been used by both HCI~\cite{simard2017machine, Hong2020Crowdsourcing, sanchez2022deep} and ML~\cite{zhu2015machine, zhu2018an} communities with different definitions.
To avoid conflicts, we utilized the term of Interactive Machine Teaching defined by Ramos~\etal~\cite{ramos2020imt}: IMT is \textit{``an IML (interactive machine learning) process in which the human-in-the-loop takes the role of a teacher, and their goal is to create a machine-learned model''}. 
This definition emphasizes user experience rather than mathematical challenges, and is well aligned with the scope of this work.

IMT systems were typically designed for non-ML-experts to build their own ML models without requiring technical knowledge and skills~\cite{dudley2018review, fails2003interactive}. 
Existing work conducted qualitative studies with ML novice users and presented user requirements and design guidelines for IMT systems~\cite{Gil2019Towards, Sanchez2021how, sanchez2022deep}. 
For example, Fiebrink~\etal's work~\cite{Fiebrink2011Human} suggested informing users of \textit{``where and how the model was likely to make mistakes''} so that users can systematically assess the benefits and risks in their applications.
Yin~\etal~\cite{Yin2019Understand} found that non-ML-experts users evaluated the model only based on its accuracy, and they were not often aware of the potential unreliability of the model when it was used in another application.



Zhou and Yatani~\cite{zhou2021enhancing} enhanced the model assessment process by visualizing the image regions that were highly weighed for predictions. 
They further found that simple teaching without further fine-grained annotations~\cite{Carney2020teachable} could cause unexpected failures. 
Effective approaches for specifying the portions of the image in the teaching process are still under-explored.

This work introduces a \vimt system that exploits users' deictic gestures to present objects for identifying the region which a model should focus on for learning.
This interface design can solve the issue of accidental use of unrelated visual features by ML models by exploiting interactions people would normally perform during the teaching phase.

\subsection{Interactive Annotations}
One standard approach to addressing accidental use of unrelated visual features is to provide annotations (e.g., a segmentation mask over the target object) and inform a system of where a model should focus.
While offering useful information, annotation is generally a tedious manual task.
Drawing a polygon-based contour on an object~\cite{boris2020cvat} is a common approach to generating a segmentation mask, but this is generally very time-consuming.
By incorporating computer vision methods, research has demonstrated different ways to reduce input from users~\cite{Laielli2019LabelAR}, including clicks~\cite{mortensen1998interactive, fbrs2020sofiiuk}, sketches~\cite{rother2004grabcut, Zhang2020weakly}, and mouse drags~\cite{chang2021spatial}. 

As these annotation tools are not specifically designed for the integration into \vimt systems, users would have to use them in a post-hoc manner.
This does not thus fully exploit the user interaction that occurred in the teaching phase for inferring segmentation masks on target objects.
Instead of proposing another annotation approach, our work utilizes users' deictic gestures toward target objects when they are performing demonstrations to a camera.
In this manner, our system achieves in-situ object annotations while teaching in \vimt systems.

\subsection{Interactions Using Deictic Gestures}


Prior research found that infants already have an ability to perform and interpret hand gestures~\cite{caselli1990communicative, meltzoff1995understanding}.
Inspired by this inherent human capability, HCI research has developed various interfaces using deictic gestures~\cite{Wobbrock2009user-defined, Vuletic2019systematic}.
One of the earliest work in this space is ``Put-that-there''~\cite{Bolt1980put}, in which users can manipulate virtual objects through a combination of deictic gestures and natural languages.
Interface applications of deictic gestures also include drone manipulations ~\cite{cauchard2015drone}, Human-Robot Interaction (HRI)~\cite{sauppe2014robdeic, Pizzuto2019Exploring} and commutations in Mixed Reality~\cite{cai2018gesture}.
Sauppe~\etal~\cite{sauppe2014robdeic} demonstrated a human-like robot that can perform deictic gestures, and found that these gestures can contribute to improving communicative accuracy in interactions with users.

Besides deictic gestures, research has investigated different aspects of hand-object interactions.
By assuming that the object under humans' manipulations would follow 1-DOF movements, Hartanto~\etal~\cite{Hartanto2020Hand} created a method to segment the object and classify the type of object motions (pure displacement motion by the prismatic joint or pure rotational motion by the revolute joints).
Other work built datasets of hand-object interactions~\cite{Cao2021reconstructing, Shan2020DaysofHands, Lee2019Hands}, and aimed to derive data-driven approaches for recognizing these interactions.
\major{Lee and Kacorri~{\cite{Lee2019Hands}} created the TEgO dataset and a system for people with visual impairments to recognize a pre-defined set of daily-life objects and assist interaction with them.}

This work extends the application of deictic gestures to \vimt systems and allows users to perform in-situ annotations while teaching in real time.
\major{More importantly, LookHere advances the generalizability by removing those constraints in prior work (e.g., pre-defined object categories~{\cite{Lee2019Hands}} or specific motions associated with holding~{\cite{Hartanto2020Hand}}).}

\section{Research Challenges and Questions}
\subsection{Challenges in Existing \vimt Systems}
\label{sec: challenges}
After reviewing the existing \vimt systems, the authors summarized our perceived challenges in the following two aspects:
\setlength{\leftmargini}{20pt}
\begin{itemize}
    \item[\textbf{C1.}] \textbf{ML models created through simplified processes supported by \vimt can be unreliable because they may learn features unrelated to target objects.} One major shortcoming of \vimt is that ML models created through such systems may unpredictably attend to unrelated objects, which is aligned with the findings by Zhou and Yatani's~work~\cite{zhou2021enhancing}. To simplify the creation of ML models for non-experts~\cite{dudley2018review, simard2017machine}, \vimt systems typically only ask users to perform several demonstrations to the camera~\cite{Carney2020teachable, francoise2021marcelle}. During teaching, the computer not only captures the object to be classified, but also other unrelated backgrounds or objects. 
    A model thus may consider those unrelated features as critical components of the target objects while users expect that it would only capture the features on the target objects. 
    This discrepancy could result in unexpected inaccuracy when the model is brought to actual use. 
    This can greatly degrade the usability of the created model and affect users' trust~\cite{Yin2019Understand} toward it.
    \item[\textbf{C2.}] \textbf{Post-hoc annotations can diminish the overall usability of \vimt systems.}
    A na\"{i}ve approach to solve the aforementioned issue is to ask users to specify what they want to be included in models (i.e., annotate the image regions of the objects). 
    Existing work has successfully simplified data annotations~\cite{sofiiuk2021reviving, mortensen1998interactive}, but these interfaces are mostly designed for more professional use~\cite{kasten2021layered}.
    Furthermore, creating a reliable ML model usually requires the user to provide many samples per class.
    Performing annotations on many images repeatedly in addition to teaching through \vimt systems can thus be overwhelming to non-expert users~\cite{smith2020no}. This also can discourage casual use of ML, which many \vimt systems envision. 
\end{itemize}
While formative user studies could further confirm these challenges, we decided not to execute them as they are already well explained in the existing literature.
Our user evaluation results presented in Section~\ref{sec: result} also confirm these challenges well.



\subsection{Research Questions}
\label{sec: rq}
\major{This work explores approaches to solve both challenges by \textit{integrating annotations into the teaching process}.
To achieve this integration, we exploit how people interact with objects of interest using \textit{deictic gestures} when they perform demonstrations to a camera.
For example, they may hold the object with both hands or may point to the object with their index fingers.
Such human behaviors are known in prior HCI research~{\cite{Kacorri2017People}} that led to diverse gesture-based applications~\mbox{\cite{pei2022hand, Lee2019Hands}}.
Therefore, we hypothesized that such deictic gestures would be an important cue for in-situ annotation.
Accordingly, we derive the following two research questions to be answered through this research:
}

\setlength{\leftmargini}{25pt}
\begin{enumerate}
    \item[\textbf{RQ1.}] \textit{How can users' deictic gestures toward objects of interest be utilized for annotations during teaching?}
    \item[\textbf{RQ2.}] \textit{Can such in-situ annotations inferred from users' deictic gestures reduce the overall teaching workload while maintaining the model accuracy?}
\end{enumerate}

RQ1 asks for technical approaches for leveraging humans' deictic gestures for the integration and the corresponding implementation.
RQ2 investigates the efficacy of such gesture-aware annotation methods.
Our design and implementation of LookHere in the following content explore RQ1, and our evaluation study answers RQ2 using multiple metrics (i.e., time consumption, model accuracies and subjective workload).

\begin{figure*}[t]
    \centering
    \begin{subfigure}[b]{0.49\linewidth}
        \includegraphics[width=\textwidth]{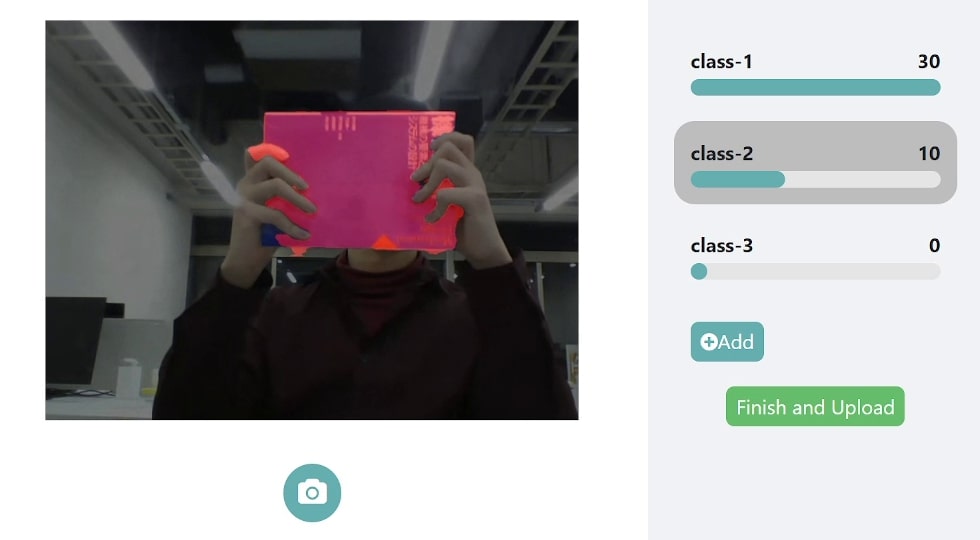}
        \caption{Teaching Interface.}
        \label{fig:teach_interface}
    \end{subfigure}
    \begin{subfigure}[b]{0.49\linewidth}
        \includegraphics[width=\textwidth]{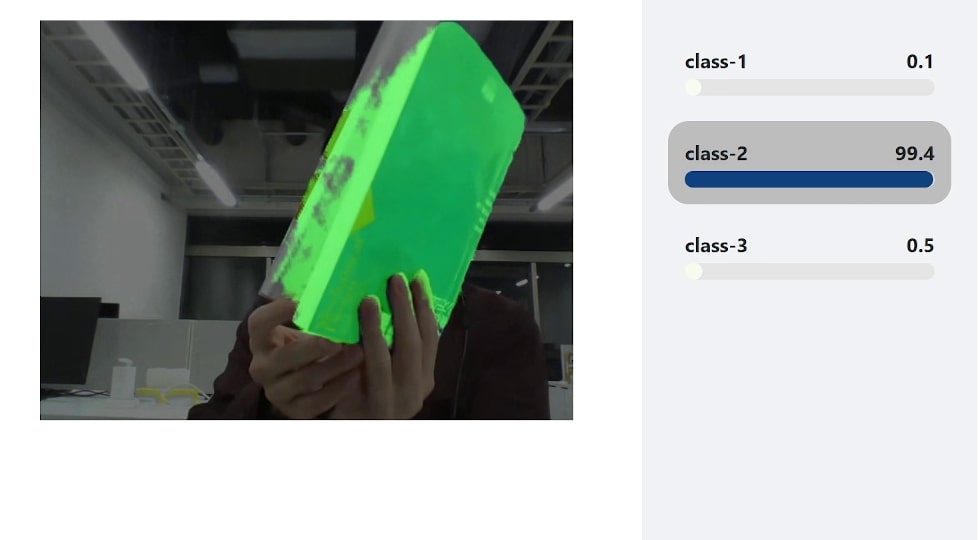}
        \caption{Model Assessment Interface.}
        \label{fig:assess_interface}
    \end{subfigure}
    \caption{The screenshots of \systemname.
    (a) In this teaching interface, real-time object highlights are provided. The number of samples per class is presented on the right side of the view;
    (b) In this model assessment interface, the saliency map visualizations for the prediction of a specified class (i.e., class 2 in this example) are shown along with the prediction confidence score. This feedback informs users of what visual features in a given frame a model is weighed for predictions.
    }
    \label{interface}
\end{figure*}

\begin{figure}[t]
    \centering
    \begin{subfigure}[b]{0.32\linewidth} 
        \centering
        \includegraphics[width=\textwidth]{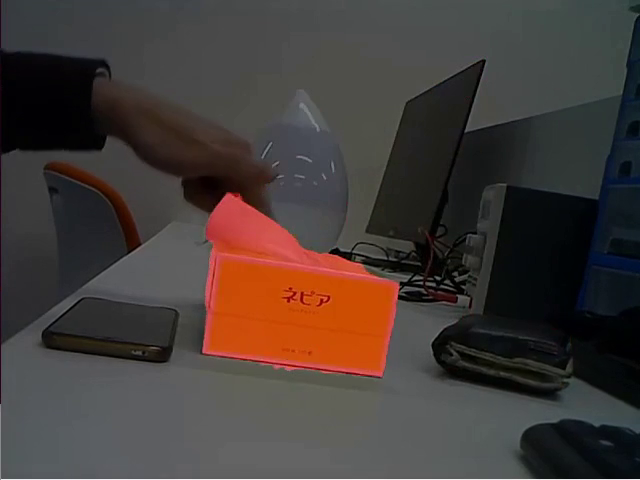}
    \end{subfigure}
    \begin{subfigure}[b]{0.32\linewidth}
        \centering
        \includegraphics[width=\textwidth]{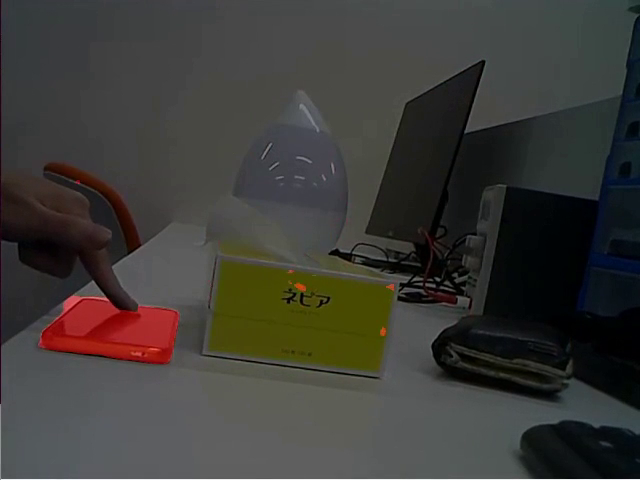}
    \end{subfigure}
        \begin{subfigure}[b]{0.32\linewidth}
        \centering
        \includegraphics[width=\textwidth]{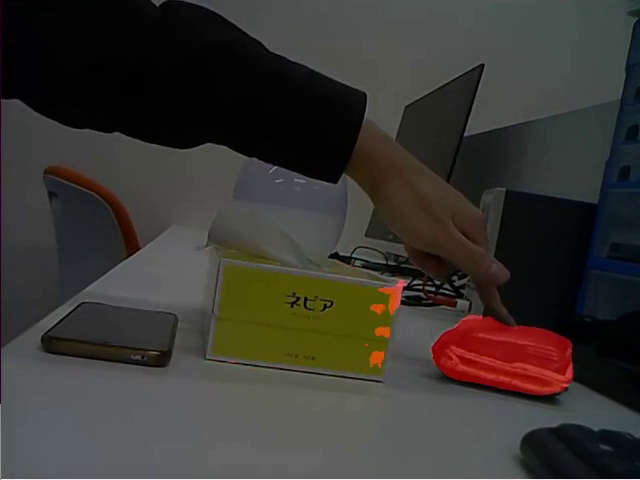}
    \end{subfigure}
    \caption{Highlights are overlaid onto different objects depending on users' deictic gestures.
    }
    \label{fig: diff-gesture-highlight} 
\end{figure}

\section{LookHere}
\label{sec: workflow}
Our \vimt system, LookHere, considers users' gestures to objects for building accurate ML models. 
Unlike existing workflows in \vimt, LookHere directly integrates the annotation process into the teaching process. 
More specifically, LookHere includes a function called \textit{object highlights} to inform which part of the camera view the system is considering as the region of the object to be learned.
In the assessment phase, LookHere also supports a model assessment process by providing a similar visualization, allowing the user to assess whether the trained model attends to the correct features.

Besides these two features explained in this section, the architecture and interaction walkthrough are similar to existing \vimt systems.
In our current implementation, users can train a multi-class classifier (i.e., classifying different objects).
To define a class, the user first selects the corresponding class (see the top-right corner of Figure~\ref{fig:teach_interface}).
Then, they can perform demonstrations of the object to the camera, and the system captures the frame when the user clicks a camera button.
The number of frames collected for each class is presented as a bar graph.
After finishing teaching for the three classes, users may either click the ``Add'' button to include more classes or the ``Finish and Upload'' button to finish the teaching session.
Appendix~\ref{apdx: model_training} shows our detailed configurations in the ML process after teaching.


\subsection{Object Highlights and In-situ Object Annotation}
\label{sec: interface_highlight}
During the teaching process, users can receive visual feedback about 
which portion of the camera view LookHere is currently considering as the region of the objects of interest.
As is shown in Figure~\ref{fig:teach_interface}, our system infers the object region based on deictic gestures users are performing (e.g., holding or pointing to an object for teaching).
Users may simply change how to perform gestures to express different target objects, as shown in Figure~\ref{fig: diff-gesture-highlight}.
LookHere incorporates a gesture-aware algorithm (see details in the next section) to achieve this adaptive highlight on objects.

Another advantage of providing this highlight in real time during the teaching process is to help users avoid including erroneous demonstrations.
Users can easily opt out of such frames by not clicking the camera button.
In this manner, LookHere takes a mixed-initiative approach~\cite{horvitz1999principles} for teaching.

When the user records the current frame by a button click, the system stores the RGB image as well as the inferred object segmentation mask.
Both data are used for model training.
In this manner, LookHere achieves in-situ object annotations during the teaching process.

\subsection{Model Assessment with Saliency Map Visualizations}
After the teaching phase, LookHere offers the model assessment mode like other \vimt systems~\cite{francoise2021marcelle, Carney2020teachable}.
However, unlike these systems, LookHere provides saliency map visualizations for users to confirm whether the created model is considering appropriate visual features. 
Figure~\ref{fig:assess_interface} illustrates an example of the view in this assessment phase.
The interface presents two visualizations for the users: bar graphs to present confidence score distributions (Figure~\ref{fig:assess_interface} right) and real-time saliency map visualizations (Figure~\ref{fig:assess_interface} left).
The confidence score shows how confidently the model considers that the current frame belongs to the corresponding class.
In the example of Figure~\ref{fig:assess_interface}, the model is $99.4\%$ confident that the object in the frame belongs to is class 2 (which is configured as a ``book'' class in Figure~\ref{fig:teach_interface}).
Real-time saliency map visualizations then help users understand which portion of the frame the model considers as the object of interest (a book in this example).
Existing work~\cite{zhang2022debiased, zhou2021enhancing} leveraged CAM methods to present such visualization while we introduce a new method for more accurate visualizations by utilizing the object segmentation masks originally generated for object highlights (see more details for Section~\ref{sec: joint_training}).


\section{Implementation}
The current prototype of LookHere is implemented as a web-based interface, and most of the computations are executed at the back end.
We use WebRTC to synchronize the video between the interface and server for real-time image processing.
The two key features presented in Section~\ref{sec: workflow} are supported by two technical components: gesture-aware object highlights and joint training.
We explain the details of the implementation of these components in this section.


\begin{figure}
\centering
\includegraphics[width=\linewidth]{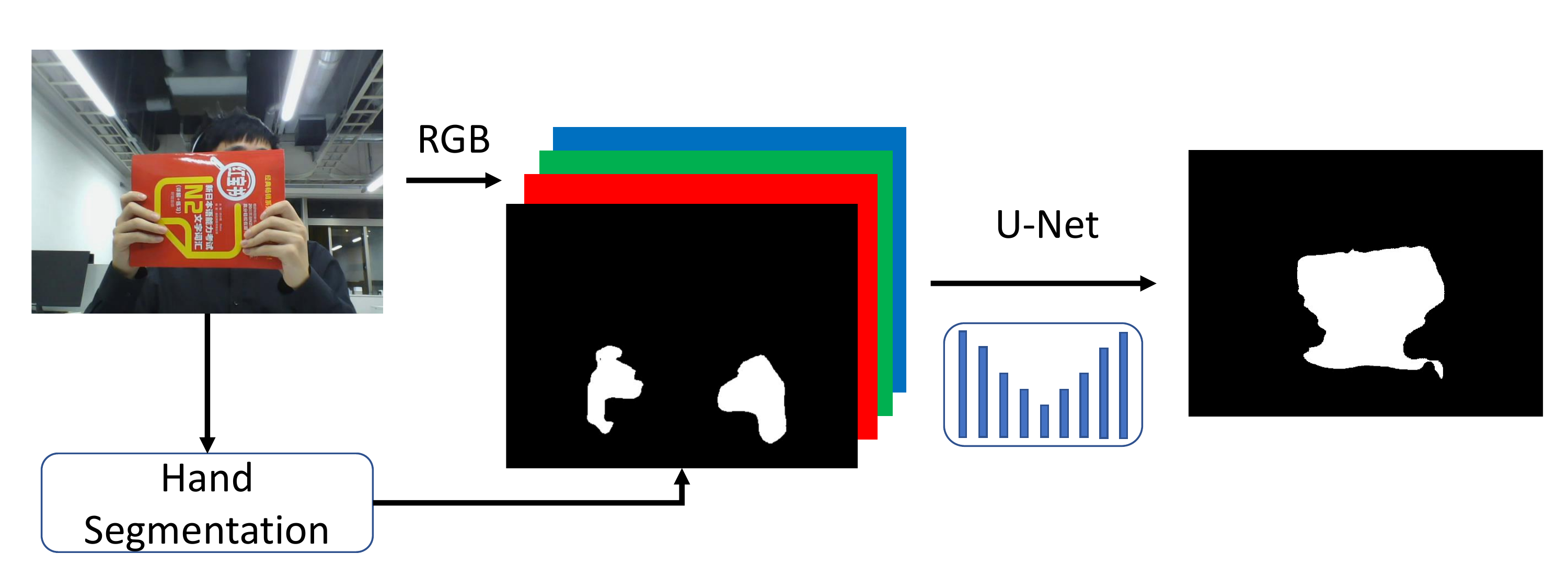}
\captionof{figure}{The generation process of object highlights.
\systemname first performs a hand segmentation with the given RGB image.
The system then feeds both the RGB image and segmentation mask into U-Net, which predicts a segmentation mask of the object guided by deictic gestures.
}
\label{fig:implementation workflow}
\end{figure}

\subsection{Gesture-aware Object Highlights}
Figure~\ref{fig:implementation workflow} summarizes the workflow of our gesture-aware object highlight algorithm.
The algorithm first applies a hand segmentor on the input image and predicts a hand segmentation mask.
It then feeds both the original RGB image and the hand segmentation mask into U-Net~\cite{ronneberger2015unet}, which outputs a segmentation mask of the object that is referred to by the users' deictic gestures.

\subsubsection{Hand Segmentation}
\label{sec: hand_seg}
We utilize Li~\etal's algorithm~\cite{Li2020self} trained on the LIP dataset~\cite{Gong2017LIP} to perform real-time hand segmentation. 
The LIP dataset parses a person into 20 body parts and garments (e.g., ``left-leg'', ``gloves'' and ``pants''), and we regard the segmentation result of ``left-arm'' and ``right-arm'' as the portion of hands.
We note that the definition of ``arm'' in the LIP dataset includes both arms and hands that are not covered by clothes or gloves. 
We notice that the publicly-available model provided by the authors of the LIP dataset is not suitable because it utilizes resnet-101 backbone~\cite{He2016resnet}, which is a very deep CNN architecture and is not executable in real time.
Therefore, we re-design their methods based on resnet-18, a much lighter model with the same encoding approach.
\major{We then tested this light-weighted model on the LIP dataset.
The result $mIoU$ accuracy of the light model is $0.621$, and that of the original model using resnet-101 is $0.680$.
This demonstrates that our light model for real-time uses can still achieve comparable accuracy to the original deep model.
}

\subsubsection{Object Highlights}

As explained in Section~\ref{sec: interface_highlight}, object highlights offer immediate feedback on what portions of the image frame the model to be trained should focus on.
To avoid losing the generalizability of \vimt, LookHere should be able to segment the object of interest in an object-agnostic manner.
To tackle this challenge, we feed the RGB image concatenated with the hand segmentation mask inferred from the hand segmentor (Section~\ref{sec: hand_seg}) into a recognition model as shown in Figure~\ref{fig:implementation workflow}.
Intuitively, this hand segmentation mask carries the information of what objects in the frame users are specifically referring to in their demonstrations.



The current implementation uses U-Net~\cite{ronneberger2015unet} as the encoder-decoder architecture.
It performs the best as well as the fastest among four commonly-used segmentation model architectures (see Appendix~\ref{apdx: arch_sel} for detailed data).
The network uses the EfficientNet~\cite{tan2019efficientnet} backbone, a design toward high computation efficiency.
To train this U-Net, we use our own dataset which we will explain in Section \ref{sec: dataset}.

\subsection{Joint Classification and Segmentation for Saliency Map Visualizations}
\label{sec: joint_training}
\begin{figure}[t]
    \centering
    \begin{subfigure}[b]{0.32\linewidth}
        \includegraphics[width=\textwidth]{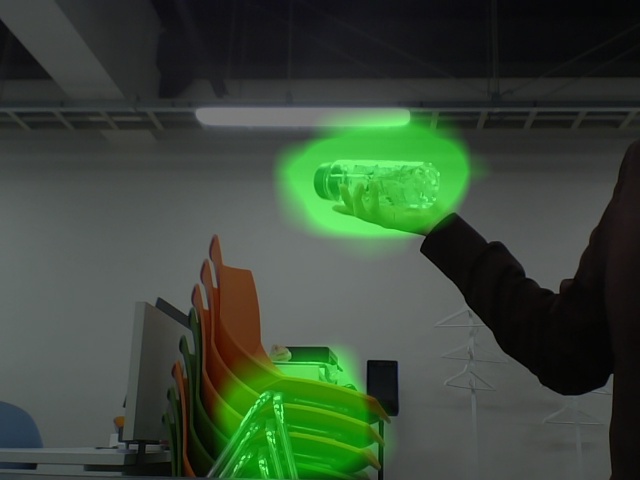}
        \caption{CAM ($\Lambda=0$).}
        \label{fig: lambda_cam}
    \end{subfigure}
    \begin{subfigure}[b]{0.32\linewidth}
        \includegraphics[width=\textwidth]{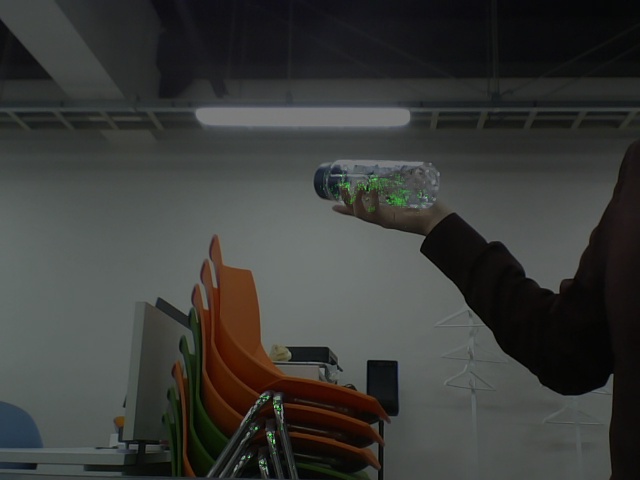}
        \caption{$\Lambda=1$.}
        \label{fig: without_aug}
    \end{subfigure}
    \begin{subfigure}[b]{0.32\linewidth}
        \includegraphics[width=\textwidth]{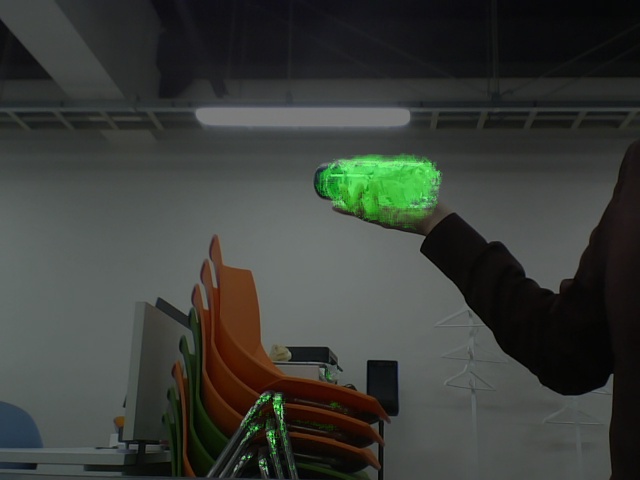}
        \caption{$\Lambda=0.718$}
        \label{fig: with_aug}
    \end{subfigure}
    \caption{Visual comparison of saliency maps with different settings of $\Lambda$. 
    The parameter $\Lambda$ in Equ.~\ref{equ: Lambda} controls the weight balance between the results by CAM and our trained model.
    }
    \label{fig: aug}
\end{figure}

Saliency map visualizations are useful for users to understand what specific portions of a given image are weighed more in their ML models.
Existing work~\cite{zhou2021enhancing, zhang2022debiased} created saliency maps of a classification model through CAM methods~\cite{zhou2016cam, Selvaraju2017gradcam}.
CAM methods are primarily used for simple classification models trained by the dataset without segmentation masks. 
Unlike existing \vimt systems, our training data accompany the object segmentation masks inferred during the teaching phase.
We thus devise a new model training approach for LookHere to exploit this unique information resource to achieve more accurate saliency maps.

LookHere identifies the areas to be highlighted by saliency maps through solving a classification and segmentation problem jointly.
This means that our backend model predicts a class as well as infers the segmentation of the object of interest at the same time.
More specifically, we train the model through a joint loss function ($l_{joint}$), which is a weighted sum of classification loss ($l_{cls}$) and segmentation loss ($l_{seg}$): $l_{joint} = l_{cls} + \lambda \cdot l_{seg}$.
$\lambda$ is a trade-off weight that determines the relative importance between the classification loss and segmentation loss.
In our current prototype, we set $\lambda$ to 1, making both of them equally important in the training process. 

While the segmentation masks originally created for object highlights can be useful for training our backend model for saliency maps as we discussed above, they may also contain some errors because the generated mask is not always perfect.
Such errors may lead to degradation in the accuracy of segmentation inference for saliency maps.
To eliminate this effect, we introduce another parameter ($\Lambda$) to control the balance between the inference results by our backend model and CAM methods:

\vspace*{-3.5mm}
\begin{equation}
\label{equ: Lambda}
    \Lambda \cdot Out + (1-\Lambda) \cdot CAM
\end{equation}

$Out$ represents the segmentation output of the our backend model and $CAM$ is the CAM inference result.
A larger $\Lambda$ value means that the system weighs more on our inference result for the output for saliency maps.

We found that taking such trade-off in consideration can greatly improve the accuracy of our saliency maps in some challenging cases.
Figure~\ref{fig: aug} illustrates the effect of $\Lambda$ in a case where a user is holding a plastic bottle.
The saliency map visualization can be quite erroneous when we only use the results of CAM (Figure~\ref{fig: lambda_cam}).
This approach would include regions that are not related to the object of interest.
On the other hand, when we only use the prediction by our backend model, the result tends to be overly conservative (Figure~\ref{fig: without_aug}).
One reason of this issue is over-fitting.
In this example, we deliberately used different backgrounds for training and testing.
As the bottle in this example was transparent, the model might have included (or overfit) some visual features of the background during training.
Such features would not appear when the background was changed when being tested, and this could thus explain why our model can be very conservative.

By choosing an appropriate value for $\Lambda$, the saliency map can visualizes the object region more precisely (Figure~\ref{fig: with_aug}).
We chose $\Lambda$ value to be the accuracy of our object highlights in our current implementation and technical evaluation (i.e., 0.718 using EfficientNet-b0 backbone).
It is out of our scope to investigate how to achieve optimization on this parameter.




\section{Deictic Gesture Dataset}
\label{sec: dataset}
\begin{figure*}[t]
    \centering
    \begin{subfigure}[b]{0.495\linewidth} 
        \centering
        \includegraphics[width=\textwidth]{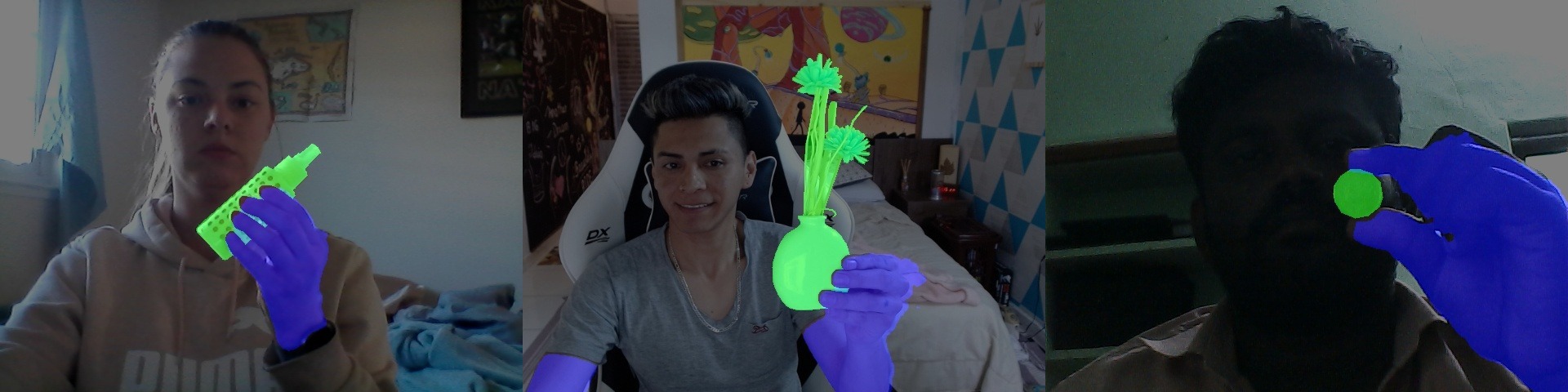}
    \end{subfigure}
    \begin{subfigure}[b]{0.495\linewidth}
        \centering
        \includegraphics[width=\textwidth]{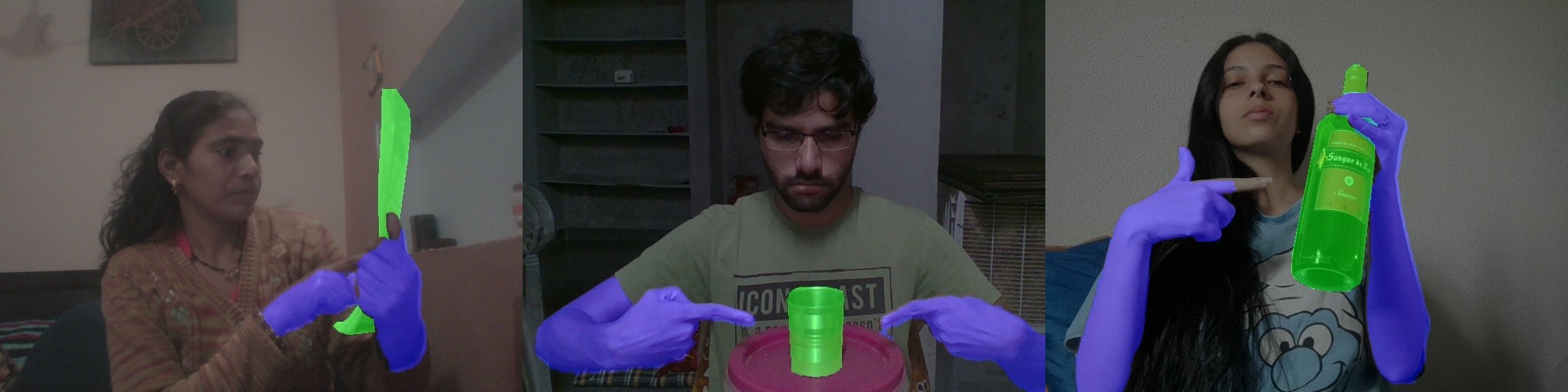}
    \end{subfigure}
    \begin{subfigure}[b]{0.495\linewidth} 
        \centering
        \includegraphics[width=\textwidth]{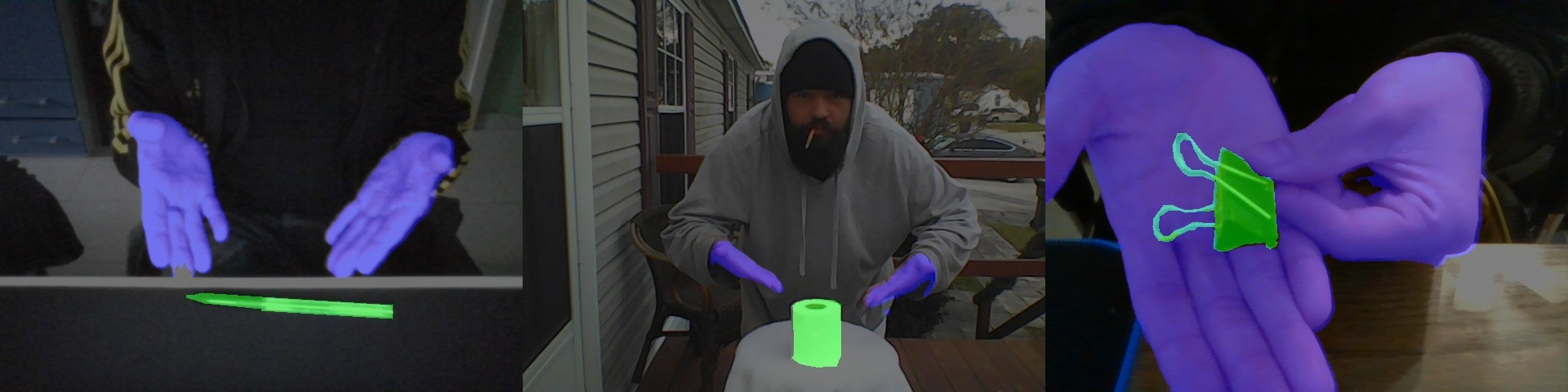}
    \end{subfigure}
    \begin{subfigure}[b]{0.495\linewidth}
        \centering
        \includegraphics[width=\textwidth]{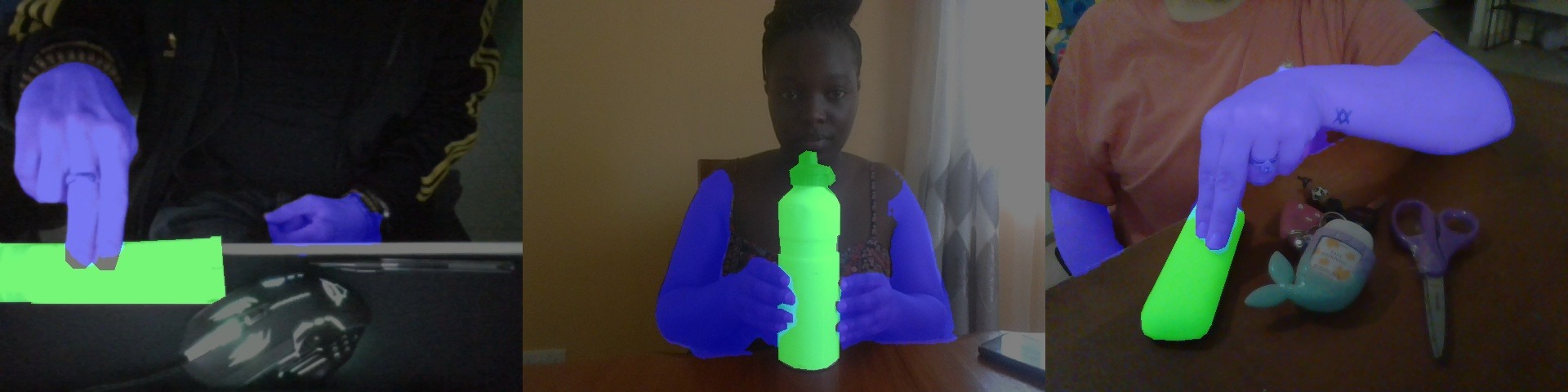}
    \end{subfigure}
    \caption{Example images in \datasetname dataset. 
    \datasetname covers four kinds of deictic gestures to objects: exhibiting (top-left), pointing (top-right), presenting (bottom-left) and touching (bottom-right).
    The hands and objects of interest are highlighted in blue and green, respectively.
    }
    \label{fig: hutics} 
\end{figure*}


\subsection{\major{Motivation of Data Collection}}
\label{sec: data_limit}
As explained in the previous section, the backend model for object highlights needs training data of how people perform deictic gestures to objects to a camera.
Among existing related human-object datasets~\cite{Damen2018EPICKITCHENS, Damen2021RESCALING, Shan2020DaysofHands, Lee2019Hands}, TEgO~\cite{Lee2019Hands} is the one that best fits our task.
TEgO includes 5758 labeled egocentric images of hand-object interactions.
For each image, there is a hand segmentation mask and a point-level annotation of the object location, which is not immediately sufficient for our purpose (object segmentation).
We therefore attempted to infer the segmentation mask of the object by emulating a click-based interactive segmentation method~\cite{sofiiuk2021reviving}.
We then manually inspected all the generated results and removed data samples where the inferred segmentation masks were completely inaccurate.
This constitutes our customized dataset with automatically-synthesized object segmentation masks, called TEgO-Syn ($n$=5232).

The trained network using TEgO-Syn achieved $mIoU$=0.895 on the testing set, showing a seemly-promising result.
Appendix~\ref{apdx: configure} provides our detailed training configurations.
To further evaluate the robustness of the network in real applications, we experimented with this model with images where various objects were presented through different deictic gestures.
Our observations showed that the model was not robust enough which we will further confirm in Section~\ref{sec: techeval}.
We then summarized three main reasons why TEgO still cannot fit our target task:
\setlength{\leftmargini}{10pt}
\begin{itemize}
    \item \textbf{A limited set of gestures}. All data in TEgO were collected from two participants, which is insufficient to cover how different people interact with the object using gestures.
    \item \textbf{A limited set of objects}. TEgO-Syn includes 5232 images of 19 objects. Training on a small set of objects repeatedly enables the model to over-fit the features of these specific objects, which is harmful to our target task, i.e., object-agnostic segmentation.
    \item \textbf{Egocentric images}. The images in the TEgO dataset are taken from the egocentric view. Our system uses a front-facing camera, which is a common configuration in \vimt~\cite{Carney2020teachable}.
\end{itemize}

\begin{figure}[t]
    \centering
    \begin{subfigure}[b]{\linewidth}
        \begin{minipage}{0.48\linewidth}
        \includegraphics[trim={0 0 23cm 0},clip, width=\textwidth]{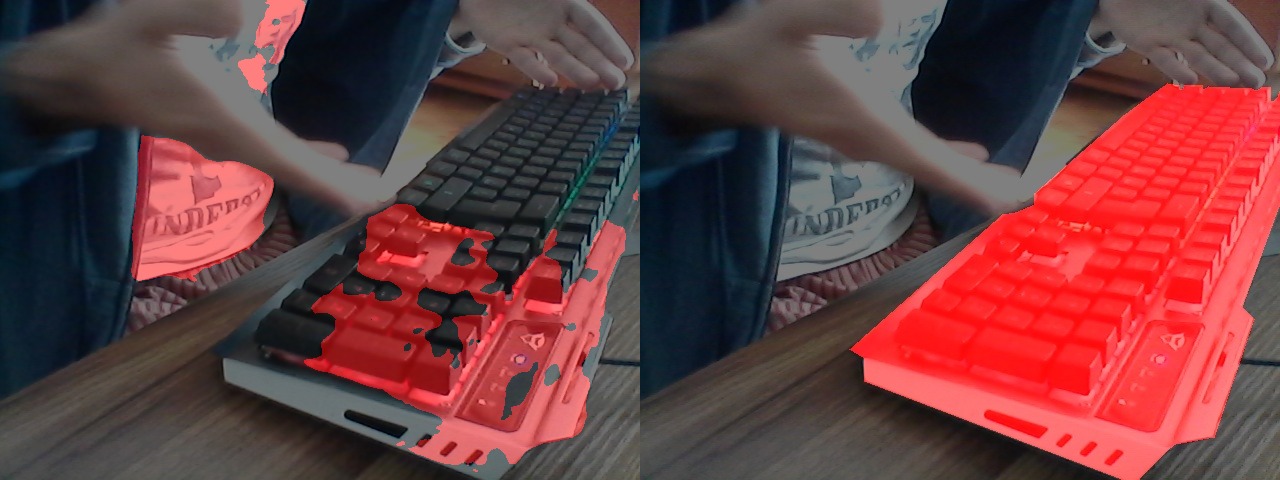}
        \vspace*{-5mm}
        \caption*{Prediction.}
        \end{minipage}
        \begin{minipage}{0.48\linewidth}
        \includegraphics[trim={23cm 0 0 0},clip, width=\textwidth]{figures/dataset/IoU0366.jpg}
        \vspace*{-5mm}
        \caption*{Ground truth.}
        \end{minipage}
        \vspace*{-2mm}
        \caption{An example with the model trained with TEgO-Syn ($IoU$=0.366).}
        \label{fig: vis_example_tego}
    \end{subfigure}

    \begin{subfigure}[b]{\linewidth}
        \begin{minipage}{0.48\linewidth}
        \includegraphics[trim={0 0 23cm 0},clip, width=\textwidth]{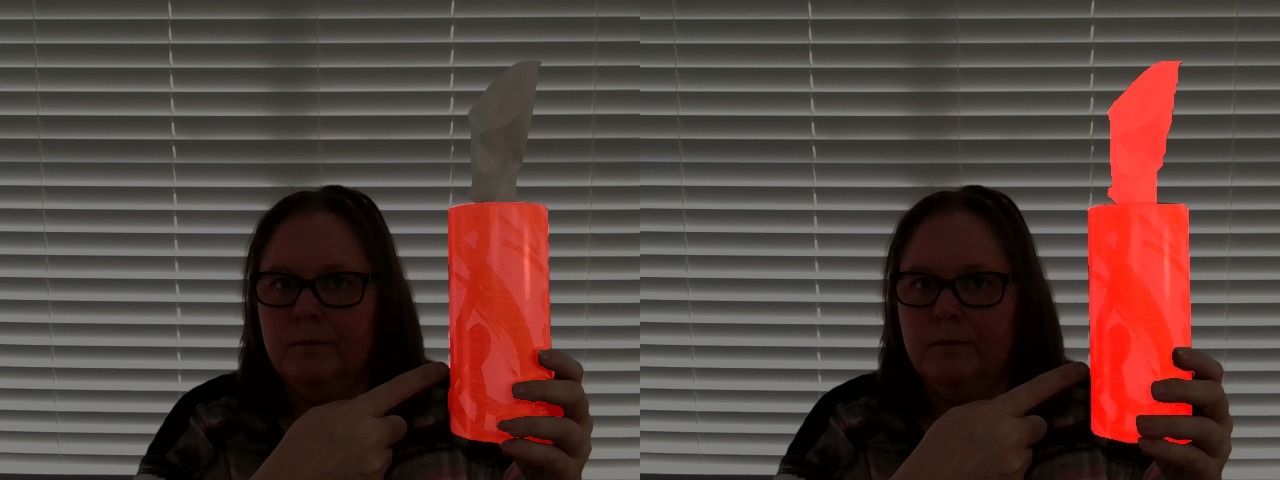}
        \vspace*{-5mm}
        \caption*{Prediction.}
        \end{minipage}
        \begin{minipage}{0.48\linewidth}
        \includegraphics[trim={23cm 0 0 0},clip, width=\textwidth]{figures/dataset/IoU0719.jpg}
        \vspace*{-5mm}
        \caption*{Ground truth.}
        \end{minipage}
        \vspace*{-2mm}
        \caption{An example with the model trained with~\datasetname~($IoU$=0.719).}
        \label{fig: vis_example_hutic}
    \end{subfigure}

    \caption{Visual comparison of predictions by the models trained with the two datasets (TEgO-Syn and \datasetname).}
    \label{fig: vis_example}
\end{figure}


\subsection{\datasetname Dataset}


To address the three issues above, we created our own dataset.
We recruited crowd-workers on Amazon Mechanical Turk, aiming to enhance the diversity of the dataset.\footnote{We received IRB approval for this data collection at our university.}
In each task, the worker needed to upload 12 images in total that clearly showed how they would use deictic gestures to express the references to objects.
For collecting a diverse set of images from each worker, we first classified deictic gestures into four categories based on Sauppe~\etal's taxonomy~\cite{sauppe2014robdeic}: pointing, presenting, touching and exhibiting.
We then asked the workers to take three different photos for each gesture category.
We also provided example pictures to clarify our expectations to the workers.

We collected 2040 qualified images from 170 crowd-workers (M: 99; F: 71) in total.
The average age of the workers was 34 ($SD$: 9.2).
Example unqualified submissions included images that were highly blurry or where no gesture was involved at all.
The crowd-workers spent 15 minutes on average to complete the task, and we paid each participant 2 dollars.
We then recruited another five people on our local crowdsourcing platform to annotate object segmentation masks on the collected images.
On average, each annotation worker labeled 408 images, and we compensated them with approximately 78 dollars on average in our local currency.
During the annotation, the workers used AnnoFab~\cite{annofab}, an online polygon-based tool, to label the segmentation masks.

Figure~\ref{fig: hutics} presents example images with the annotated object segmentation masks.
Unlike TEgO, our dataset contains a wide range of objects, deictic gestures, backgrounds, and environmental conditions.
Table~\ref{tab: dataset_compare} summarizes a comparison of~\datasetname~v.s. TEgO-Syn.

\begin{table*}[]
\caption{Comparison of \datasetname and TEgO-Syn.}
\label{tab: dataset_compare}
\begin{tabular}{ccccccc}
\hline
         & \# of Participants & \major{Object Types} & View         & \# of Images & Object Annotation                                                  & Target Task                                                                                      \\
      \hline
Hutics   & 170                & \major{Uncontrolled}           & Front-facing & 2040         & \begin{tabular}[c]{@{}c@{}}Segmentation\\ mask\end{tabular} & \begin{tabular}[c]{@{}c@{}}Object-agnostic segmentation\\ specified by gestures\end{tabular}     \\
\hline
TEgO-Syn & 2                  & \major{Controlled}     & Egocentric   & 5232         & Point-based                                                 & \begin{tabular}[c]{@{}c@{}}Object recognition for \\ people with visual impairments\end{tabular} \\
\hline
\end{tabular}
\end{table*}

\subsection{Performance of Object Highlights on \datasetname}
\label{sec: techeval}
We used the data from $80\%$ of the participants in \datasetname (\ie, 1632 images from 136 people) for training and $20\%$ for testing.
We trained our algorithm using the same configuration above, and the network achieves $mIoU$=0.718 and 0.806 using EfficientNet-b0 and EfficientNet-b3 backbone on the testing set, respectively.
Running on one GTX 2080Ti GPU, our implementation of the algorithm was able to reach 28.3 fps and 24.0 fps with the EfficientNet-b0 and EfficientNet-b3 backbone, respectively.
For comparison, we trained another model with the same network architecture using TEgO-Syn and tested with images in \datasetname.
The accuracy of that model was $mIoU$=0.368, much lower than that of the same network using~\datasetname for training.
This significant accuracy drop from 0.895 (tested on TEgO-Syn) further confirms our observations discussed in Section~\ref{sec: data_limit}.

Figure~\ref{fig: vis_example} shows a visual comparison of the results between the networks trained on TEgO-Syn and \datasetname.
Each example in Figure~\ref{fig: vis_example} are the one in our testing set that has the closest IoU values (0.366 and 0.719) to the corresponding mean IoU values (0.368 and 0.718).
We therefore use the model trained on \datasetname in our current prototype implementation.

\section{User Study}
\label{sec: user_eval}


We conducted a comparative user study to evaluate how LookHere could improve the experience of \vimt in terms of time cost for teaching, accuracy performance on models created, and subjective user workload.

\subsection{Interface Conditions}
\label{sec: inter_cond}
Besides LookHere, we included the following three interface conditions to represent existing \vimt and object annotation methods.

\setlength{\leftmargini}{10pt}
\begin{itemize}
\item \textit{Na\"{i}veIMT}:
This represents the most common design in current \vimt systems~\cite{Carney2020teachable, francoise2021marcelle}. In this condition, participants would only perform object demonstrations during the teaching phase.
Participants would not have an opportunity to specify which regions of the recorded images would represent the object for a given class. We implemented this na\"{i}ve IMT system based on the source code of Zhou and Yatani~\cite{zhou2021enhancing} available online. 
\item \textit{Contour}:
In addition to the teaching process with the na\"{i}ve IMT system, this condition would involve a manual annotation procedure in a post-hoc manner.
In this condition, participants would be asked to perform contour-based annotations.
This annotation style is widely used in IMT systems for medical purposes~\cite{bounias2021interactive}.
We used AnnoFab~\cite{annofab} for post-hoc contour-based annotations in this study.

\item \textit{Click}: 
The third reference condition included a click-based annotation method~\cite{sofiiuk2021reviving}.
We decided to include this condition as the annotation process would be more lightweight than a contour-based approach.
We used RITM~\cite{sofiiuk2021reviving} as the click-based annotation tool.
\end{itemize}

All these three reference conditions involve the teaching process using the na\"{i}ve IMT system.
To shorten the overall study time, we decided to ask participants to perform teaching under the two conditions of \textit{Na\"{i}veIMT} and LookHere.
After this teaching task, participants were then asked to perform annotations under the two conditions of \textit{Contour} and \textit{Click} using the data recorded under the \textit{Na\"{i}veIMT} condition.
In this manner, we liberated the participants from performing the same tasks repeatedly with \textit{Na\"{i}veIMT} for the \textit{Contour} and \textit{Click} conditions.

We counter-balanced both the condition order of \textit{Na\"{i}veIMT} and LookHere and that of \textit{Contour} and \textit{Click} across participants.
The order of tasks of teaching and annotation was fixed (the teaching process was the first).




\subsection{Evaluation Metrics}

\subsubsection{Teaching and Annotation Time}
We measured how long it took for participants to finish the model creation process under each interface condition.
Specifically, we recorded the teaching/annotation time from when the participants started uploading/annotating the first sample to when they finished the last (30th) sample.
    
\subsubsection{Model Accuracy}

\major{We measured both classification accuracy and segmentation accuracy (i.e., mean Intersection over Union or $mIoU$) of the created models.
We randomly used $80\%$ of the data for training and the rest for testing.
For classification accuracy, we utilized cross-condition validation.
Specifically, we tested three models trained by data collected from \textit{nai\"veIMT} on the data collected from LookHere, and vice versa.
In terms of the object segmentation accuracy, we only performed cross validation for data collected from LookHere because there were no ground-truth segmentation annotations in LookHere to validate models created from \textit{nai\"veIMT}.
This ensured that LookHere gained no advantage over the three comparative conditions.
}
\subsubsection{NASA-TLX} NASA Task Load Index (TLX)~\cite{hart1988development} is a standard metric for perceived workload.
We included this to understand how different conditions could affect the experience of creating ML models with different configurations of IMT systems.

\subsection{Procedure}

At the beginning of the study, we told participants that their goal was to create four AI models to classify and detect objects by the given systems.
For the teaching tasks, we allowed participants to use any object available in our experimental space.
They were also welcome to bring their own belongings in the study.
We did not limit the set of objects to be used in this experiment in order not to lose the validity of the study.
After they had explanations about the two teaching methods (\textit{Na\"{i}veIMT} and LookHere) and became comfortable with using both, they were asked to create three classes for classification, and generate 30 images for each class through the given teaching method.
After completing teaching with the two methods, participants were given an opportunity to take a break.

We next moved to the annotation tasks with the two interfaces (\textit{Contour} and \textit{Click}).
We provided explanations about these two tools, and participants were given practice time to become comfortable with using them. 
Participants were then asked to annotate all the images they captured under the \textit{Na\"{i}veIMT} condition.
They were instructed to perform each annotation task as fast and accurately as possible.
Participants were allowed to take a break between the two sets of tasks (i.e., using the two annotation tools). 

Participants were asked to fill in NASA-TLX questionnaires after finishing each of the two task sessions (teaching and annotation).
In this manner, we ensured that participants remembered their experience of the conditions.
When participants were rating NASA-TLX for \textit{Contour} and \textit{Click}, we explicitly asked them to consider their overall workload for the combination of the teaching method with \textit{Na\"{i}veIMT} and the given annotation approach.


After completing both the teaching and annotation tasks, we conducted semi-structured interviews with participants.
We first interviewed them about their overall experience and perceived benefits and shortcomings of the methods used in the study.
This helped us collect their immediate use experience of each condition without being biased by the performance of the resulting models (i.e., the accuracy of the created models).
We next offered our model assessment interface (Figure~\ref{fig:assess_interface}) for all the four resulting models.
Participants were allowed to freely use them and check whether their created model would function accurately.
We then interviewed them about how they perceived their four models and would characterize them differently.



The whole study takes approximately 3.5 hours on average.
We offered each participant compensation of approximately 40 USD in a local currency at the end of the experiment.

\subsection{Apparatus}
We set the video frame rate to be 24 fps across all the conditions.
We used the EfficientNet-b0 backbone in our object highlights (i.e., the lightest model), aiming to understand the effectiveness of such feedback even under the least accurate setting.

\subsection{Participants}
We recruited 12 non-expert participants (P1 -- P12) for this study.
None of them had experience in studying or working in the fields related to AI or ML.
Eight of them were female, and the rest were male.
The age of participants ranged from 23 to 28.

\begin{table}[t]
\caption{The mean values and standard deviations of the task completion time and accuracies (classification and segmentation) across the four interface conditions.
}
\label{tab: res_time_acc}
\begin{tabular}{cc|cc|c|c|c}
\hline
\multicolumn{2}{c|}{\multirow{2}{*}{}} & \multicolumn{2}{c|}{LookHere}     & \multirow{2}{*}{\textit{Na\"{i}veIMT}} & \multirow{2}{*}{\textit{Click}} & \multirow{2}{*}{\textit{Contour}} \\ \cline{3-4}
\multicolumn{2}{c|}{}                  & \multicolumn{1}{c|}{$\Lambda$=1} & $\Lambda$=0.718 &                             &                        &                          \\ \hline
\multicolumn{2}{c|}{Time [s]} &
  \multicolumn{2}{c|}{\begin{tabular}[c]{@{}c@{}}104\\ (44)\end{tabular}} &
  \begin{tabular}[c]{@{}c@{}}67\\ (10)\end{tabular} &
  \begin{tabular}[c]{@{}c@{}}1,197 \\ (228)\end{tabular} &
  \begin{tabular}[c]{@{}c@{}}1,483\\ (407)\end{tabular} \\ \hline
\multicolumn{1}{c|}{\multirow{2}{*}{Acc.}} &
  Cls. &
  \multicolumn{2}{c|}{\begin{tabular}[c]{@{}c@{}}0.824\\ (0.158)\end{tabular}} &
  \begin{tabular}[c]{@{}c@{}}0.847\\ (0.190)\end{tabular} &
  \begin{tabular}[c]{@{}c@{}}0.880\\ (0.141)\end{tabular} &
  \begin{tabular}[c]{@{}c@{}}0.833\\ (0.159)\end{tabular} \\ \cline{2-7} 
\multicolumn{1}{c|}{} &
  Seg. &
  \multicolumn{1}{c|}{\begin{tabular}[c]{@{}c@{}}0.578\\ (0.233)\end{tabular}} &
  \begin{tabular}[c]{@{}c@{}}0.605\\ (0.153)\end{tabular} &
  \begin{tabular}[c]{@{}c@{}}0.139\\ (0.095)\end{tabular} &
  \begin{tabular}[c]{@{}c@{}}0.716\\ (0.167)\end{tabular} &
  \begin{tabular}[c]{@{}c@{}}0.732\\ (0.151)\end{tabular} \\ \hline
\end{tabular}
\end{table}

\label{sec: user_quant}
\begin{figure*}
    \centering
    \includegraphics[width=\linewidth]{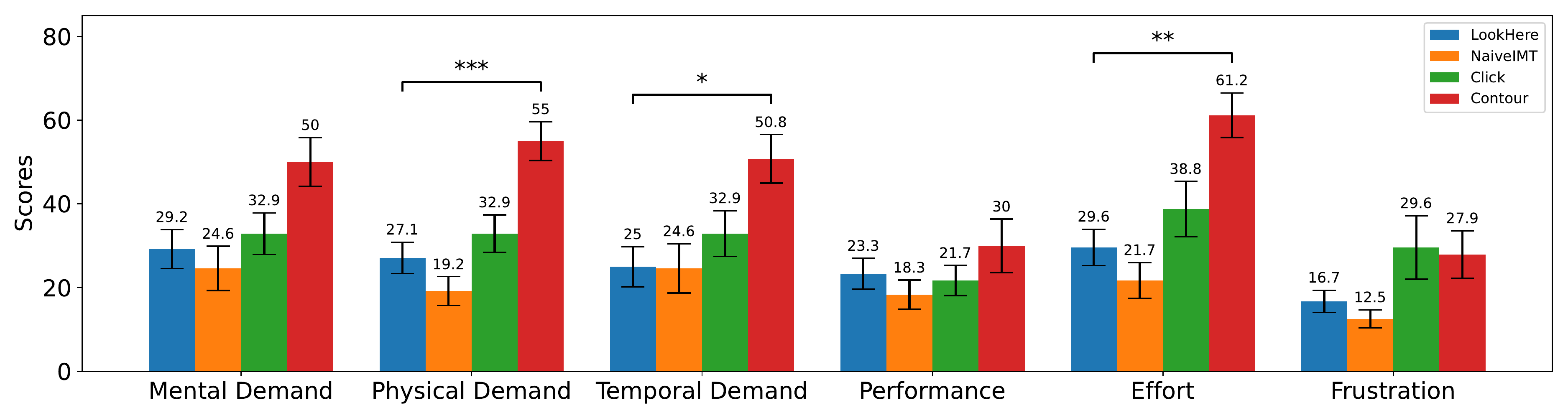}
    \caption{The results of the mean NASA-TLX scores across the six subscales. The error bars represent the standard errors. The observed significant differences are indicated with $*$, $**$ or $***$ ($p$<.05, $p$<.01 and $p$<.001, respectively).}
    \label{fig: tlx}
\end{figure*}

\section{Results}
\label{sec: result}
\subsection{Quantitative Results}
\label{sec: quant_result}


Table~\ref{tab: res_time_acc} presents the mean completion time and model accuracy in the study.
With respect to the task completion, the \textit{Contour} and \textit{Click} conditions exhibited much longer time than the other two conditions (11.5 and 14.3 times than the LookHere condition, respectively).
One-way repeated-measure ANOVA found that the factor of the conditions was significant ($F(3,33)$=134.02, $p$<.001, $\eta^2$=.92).
We then used Scheffe's multiple comparison procedure to compare the take completion time under the LookHere condition against the three reference conditions.
We found that the completion time under the \systemname condition was significantly shorter than those under the \textit{Contor} and \textit{Click} conditions (both $p<.001$).
This result clearly suggests that LookHere successfully removed the effort for object annotations.

We next looked into the accuracies of the models created under the four conditions.
As shown in Table \ref{tab: res_time_acc}, the mean accuracies for classification (predicting the correct class for the given image from the three classes defined by each participant) did not show large differences.
Our one-way repeated-measure ANOVA did not find a significant effect of the interfaces ($F(3,33)$=.460, $p$=.712, $\eta^2$=.04).

We further examined the segmentation accuracies.
As shown in Table \ref{tab: res_time_acc}, the accuracy under the \textit{Na\"{i}veIMT} condition was clearly lower than those with the other methods.
One-way repeated-measure ANOVA found a significant effect of the interface conditions ($F(3,33)$=105.98, $p$<.001, $\eta^2$=.91).
Scheffe's multiple comparison procedure revealed a significant difference between \systemname and \textit{Na\"{i}veIMT} ($p$<.001).
This result confirms that the models created with data collected under the \textit{Na\"{i}veIMT} condition did not necessarily weigh the visual features in the objects of interest, implying potential unreliability in actual use.

We also compared the segmentation predictions on the same trained model with two different $\Lambda$ values: 1 and 0.718 (the default configuration in our current implementation).
While the accuracy was improved by $0.027$ with the value of 0.718, this difference was not significant.
Future research should investigate how $\Lambda$ should be configured to achieve the best performance, but this result implies that the combination of CAM results and the inference by the backend object segmentation model could offer improvements.


We next examined the NASA-TLX results.
Figure~\ref{fig: tlx} shows the mean values of the raw NASA-TLX subscales.
One-way repeated-measure ANOVA on each subscale found significant effects by the conditions in mental demand ($F(3,33)$=7.37, $p$<.001, $\eta^2$=.40); physical demand ($F(3,33)$=16.27, $p$<.001, $\eta^2$=.60); temporal demand ($F(3,33)$=7.86, $p$<.001, $\eta^2$=.42); effort ($F(3,33)$, $p$<.001, $\eta^2$=.57); and frustration ($F(3,33)$=3.23, $p$<.05, $\eta^2$=.23).
No significant result was found in performance ($F(3,33)$=1.55, $p$=.22, $\eta^2$=.12).
Post-hoc Scheffe's procedure revealed significant differences in physical demand, temporal demand and effort between \systemname and \textit{Contour} ($p$<.001, $p$<.05, and $p$<.01, respectively).
These results suggest that the contour-based annotation method significantly impacted the user experience negatively.

In summary, the quantitative results show that LookHere was able to achieve the best balance of task completion time and model accuracy.
We further looked into how different user experience of the four interface conditions was through our qualitative results.

\subsection{Qualitative Results}
We transcribed the interviews and extracted quotes that were related to user experience and opinions about the four interfaces tested.
We then performed the open coding approach to categorize the quotes and derive them in a bottom-up manner.


\subsubsection{Burden for post-hoc annotations}
Six participants (P1, P4, P5, P9, P10 and P12) explicitly mentioned that post-hoc annotations were tedious and reduced the perceived usability of an overall \vimt system.
While nine participants preferred the \textit{Click} annotation method to \textit{Contour}, all participants agreed that both approaches were  \textit{``time-consuming''}.
\myquote{[A good process] shouldn't contain the annotation process because it is the most time-consuming one and requires lots of effort.
On the contrary, these two (\textit{Na\"{i}veIMT} and LookHere) are very comfortable to use because there is only one step.}{P4}

All participants considered LookHere as \textit{``efficient''} because it does not involve explicit post-hoc annotations.
This was clear from the task completion time and NASA-TLX results, and our qualitative data were also indicative.
In particular, P1 appreciated that LookHere combined the teaching and annotation process:
\myquote{It can greatly improve the user experience in terms of not only time consumption but also the sense of satisfaction.}{P1}

Participants were also satisfied with the accuracy of their created models achieved through LookHere.
Despite the simplified teaching experience, they could not notice the accuracy difference between LookHere and \textit{Contour}.
\myquote{I prefer to use [LookHere]. First, its accuracy is good, and it's easy to use ... It is a user-friendly design, not requiring much effort and time.}
{P10}
\myquote{In terms of effort and performance, [LookHere] is definitely a cost-effective choice ... Speaking of [\textit{Contour}], it requires much effort, but its result is not that good, probably similar to [LookHere]. It makes me feel that it is not worthwhile.}{P6}


\subsubsection{Uncertainty in teaching with \textit{Na\"{i}veIMT}}
\label{sec: uncertainty}
Participants expressed their concerns about whether the model created with the \textit{Na\"{i}veIMT} approach correctly interpreted their teaching.

\myquote{Because you can't find its focus, as a user, you can't confirm whether it (the computer) understands my idea.}{P8}
\myquote{[\textit{Na\"{i}veIMT}] is very convenient to use but I am afraid that the performance would be bad.}{P3}


On the other hand, object highlights shown in LookHere offered our participants more confidence that the regions of the objects would be considered more.

\myquote{[\textit{Different from \textit{Na\"{i}veIMT}, LookHere}] is simple, and it also has visualizations. It can let users keep well informed whether the object is recognized [by the computer].}{P5}

In case object highlights were out of place, participants adjusted their deictic gestures until they were well overlaid onto the objects.
This offered a sense of control as P12 commented:

\myquote{On one hand, the procedure is simple, and on the other hand, [LookHere] itself has already drawn that pattern (object highlights). [Even though it sometimes has errors,] I can change some positions [of the object] and it can [successfully] capture this [object] ... It provides a sense of control. Unlike [\textit{{Na\"{i}veIMT}}], I do not know what it captures [within each image].}
{P12}

\subsubsection{\major{Limitations}}
\label{sec: limitation}
Participants pointed out limitations of \systemname, and some further provided suggestions on how we can improve the current prototype.
\major{For example, P11 raised an issue that users could not interact with {\systemname} using bimanual interaction since one hand is required to manipulate the mouse, clicking on the camera icon in Figure~{\ref{fig:teach_interface}}.}

\major{Additionally, P4 pointed out that there was a lack of further teaching/clarification support when object highlights fail. 
P4 further suggested that V-IMT should integrate more functions so that users can better correct object highlights in erroneous cases, rather than passively avoid teaching these samples.}
\myquote{In terms of [LookHere], is it possible to utilize the \textit{Click} function there?
For example, when I hold something, [if object highlights are erroneous at this moment,] can I tell the computer which region I want it to recognize [by clicking on the object]?
... In the current design, I can only change the position (of the object) to adjust it (the highlight), and this makes me feel quite inactive (i.e., not in good control of object highlights).}{P4}

\section{\major{Discussion}}
\major{As mentioned in Section 3.2, this work aims to (1) explore technical solutions for the integration of object annotations into the teaching process (i.e., RQ1) and (2) study the effectiveness of the solution (i.e., RQ2).
We answer RQ1 through our implementation of LookHere as explained in Section 4 and 5. 
Our evaluation results further show that LookHere can effectively reduce users' workload during the model creation process while maintaining similar model accuracies, answering RQ2.

Despite its effectiveness, we also found several drawbacks of our systems that limited user experience in practice.
In the following content, we share our insights about how future work can improve our system and how to extend our research questions to exploit more human interactions to achieve in-situ annotations in IMT.

Additionally, our dataset, HuTics, which is designed for the gesture-aware object-agnostic segmentation task, is one important contribution of this work.
We further show that our approach and dataset can also be used to support other HCI projects by demonstrating several example applications.
}

\subsection{\major{Depth-aware Object Highlights}}
\label{sec: discussion_depth}
\major{
Despite the effectiveness of object highlights to support efficient teaching, we still observed typical erroneous cases that remain to be addressed.
Figure~{\ref{fig: pred_failure}} shows an example where {$IoU$} was low (more examples can be found in Appendix~{\ref{apdx: failure}}). 
The person in this figure is pointing at an object on her head, but our algorithm incorrectly highlights the clock in the background.
Our object highlights also tend to fail in cases where a person is pointing at an object at a distance (e.g., buildings or furniture that are not close to hands). 
These failure cases are mainly caused by the current implementation that uses 2D hand segmentation features without a 3D understanding of the scene.
A future system may consider obtaining a richer set of information through 3D scene reconstruction~{\cite{dahnert2021panoptic}}, 3D hand pose estimation~\mbox{\cite{Cao2021reconstructing, huang2020hand}} from RGB images, or directly use depth cameras~{\cite{izadi2011kinect}}.
Future research should further study how to simplify the aforementioned feature extractors to be used in real time for V-IMT systems or how to use depth sensors to support V-IMT systems~\mbox{\cite{zhang2021single, zhange2019interaction}}.
}

\begin{figure}
    \centering
    \begin{subfigure}[b]{0.49\linewidth}
        \centering
        \includegraphics[width=\textwidth]{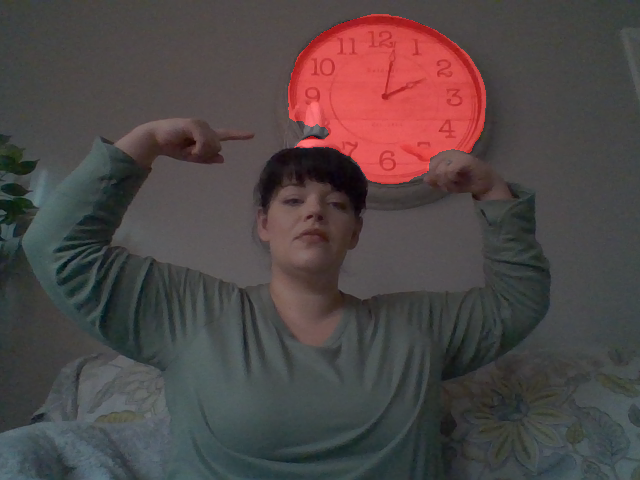}
        \caption{Prediction.}
    \end{subfigure}
    \begin{subfigure}[b]{0.49\linewidth}
        \centering
        \includegraphics[width=\textwidth]{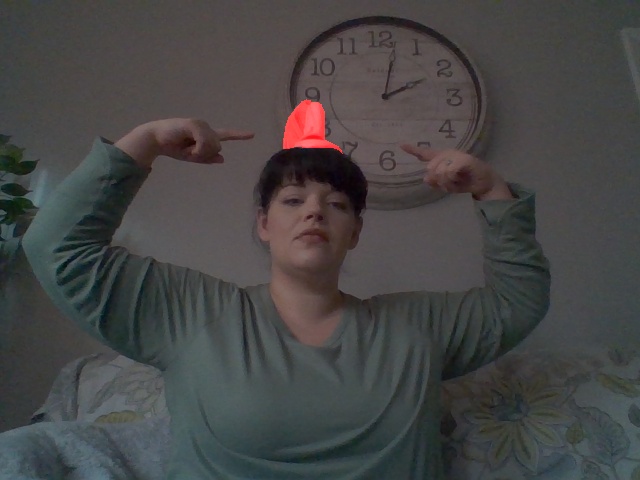}
        \caption{Ground truth.}
    \end{subfigure}
    \caption{A failure case of object highlights.}
    \label{fig: pred_failure}
\end{figure}

\subsection{\major{Voice Input and In-situ Correction}}
\major{
As mentioned in Section~{\ref{sec: limitation}}, we observed several limitations of our interface design.
To enable bimanual interactions with the object, future research can investigate how to use technologies like voice input or facial expression recognition to replace a button click. 
For example, when users want the system to sample the current frame, they can simply say ``collect'' or smile to the system while performing bimanual deictic gestures.

In addition, as mentioned in Section~{\ref{sec: limitation}}, future systems should study how to enable users to \textit{actively} correct object highlights when they observe prediction errors.
Although the segmentation annotation in LookHere allows users to choose appropriate frames for teaching, the role of users in this Human-AI collaboration is relatively passive. 
When users observe the failure case of object highlights, they should be given an opportunity to \textit{actively} correct the error~{\cite{smith2020no}}, instead of \textit{passively} avoiding those data.
Allowing such in-situ correction initiated by users can further empower the ability of IMT systems to \textit{``leverage human capabilities and human knowledge beyond labels''}~{\cite{ramos2020imt}}, achieving better human-AI collaboration.
}

\subsection{\major{Other Modalities and Privacy Issues}}
\major{
While this paper focuses on studying how to use deictic gestures to enable in-situ annotations, they are not the only human interaction that future IMT research can exploit. 
As we discussed in Section~{\ref{sec: discussion_depth}}, our gesture-aware annotation approach may not function with some deictic gestures users may perform.
More importantly, humans also innately perform other interactions as a cue of objects of interest.
For example, future research can study how to use gaze tracking technologies to capture the object of interest that is difficult to hold by hand (e.g., buildings or scenery).
While examining other modalities is out of scope of this work, future work on this aspect is encouraged.}

\major{Despite the benefits from collecting fine-grained annotation by sensing additional human interactions, such systems without proper designs may cause severe privacy issues.
We therefore encourage future research to study how to balance privacy protection and the benefits from in-situ annotations in IMT.}
\subsection{\major{Applications of the Object-agnostic Segmentation Model Trained on \datasetname}}
\major{One important contribution of this work is our object-agnostic segmentation model and its dataset, \datasetname.
Although our original objective was to enable \systemname, we envision that our model can be used for a broader range of applications, not limited to IMT research.}
\subsubsection{\major{Intelligent Virtual Background}}
\major{Using our model, developers can create an intelligent virtual background used in online meeting systems which is aware of the object users are trying to present to others.
Segmentation algorithms for virtual backgrounds do not typically consider the behavior of object presentation.
Therefore, virtual backgrounds often hide the objects held by users, diminishing user experience in certain scenarios.
Our model can address this issue and show the object held by the user while preserving virtual backgrounds to support a better communication experience.
}
\subsubsection{\major{Gesture-guided Portrait Mode}}
\major{Portrait mode in recent smartphones allows users to have a focus effect (e.g., blurring the background to highlight a person in the foreground).
Using our model, such a portrait mode may create a focus effect on the objects held by users intelligently.
Our object-agnostic segmentation model thus has a potential to enrich user experience of photo shooting with smart devices.
}
\subsubsection{\major{Supports for People with Visual Impairments}}
\major{Prior work demonstrated an assistive technology for people with visual impairments by recognizing objects held by users.
While it only recognized 19 objects constrained by the dataset used in that project, future assistive systems may develop a more generalizable approach by using our model trained on \datasetname.
They can first locate an object held by users using our object-agnostic segmentation model, and then apply a classification model trained on large-scale datasets that cover thousands of objects (e.g., 1000 classes in ImageNet~{\cite{imagenet}}), achieving the goal of recognizing various objects for supporting activities of people with visual impairments.
}
\section{Conclusion}
This work demonstrates LookHere, a \vimt system that allows users to annotate objects in real time during the teaching phase by exploiting users' deictic gestures.
We build our own dataset (\datasetname), consisting of 2040 front-facing images of deictic gestures and objects to achieve our implementation. 
Our user study results show that LookHere successfully removed substantial user effort on post-hoc manual annotations.
However, the models created through LookHere did not show significant differences in their accuracies compared to those using the data with manual annotations.

\begin{acks}
A part of this research was supported by the NII CRIS collaborative research program jointly managed by NII CRIS and LINE Corporation. Co-Design Future Society Fellowship also supports the first author of this paper.
We thank Xiang `Anthony' Chen and all the reviewers for their insightful feedback on this paper.
We also thank Anran Xu for his help in our data collection.
\end{acks}
\bibliographystyle{ACM-Reference-Format}
\bibliography{contents/reference}

\appendix

\section{Implementation Details and Extra Technical Results}
\subsection{Configurations of Machine Learning Process in \systemname}
\label{apdx: model_training}
Each image captured in the front end is fixed at the size of $480\times640$ (height $\times$ width).
We chose U-Net~\cite{ronneberger2015unet} with EfficientNet-b0 backbone~\cite{tan2019efficientnet} to be the machine learning model at the back end taught by users.
During the training stage, \systemname uses Adam optimizer and performs fine-tuning on the model pretrained on ImageNet~\cite{imagenet} for 50 epochs.
The batch size is four, and the learning rate is 1e-4.

Note that we only use the encoder of the model for the \textit{Nai\"veIMT} condition (see details in Section~\ref{sec: inter_cond})
because there is no ground truth data of segmentation masks in this condition, which is necessary for training the decoder.
We used CAM~\cite{zhou2016cam} to predict saliency maps using this classification model.



\begin{figure}[!htb]
    \centering
    \begin{subfigure}[b]{\linewidth}
        \begin{minipage}{0.48\linewidth}
        \includegraphics[width=\textwidth]{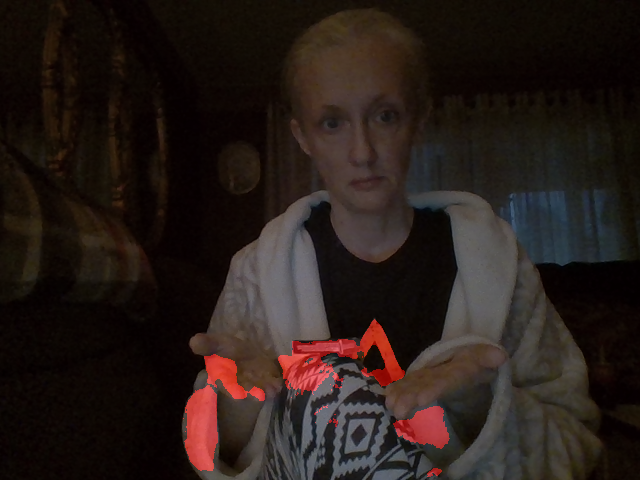}
        \caption{Prediction.}
        \end{minipage}
        \begin{minipage}{0.48\linewidth}
        \includegraphics[width=\textwidth]{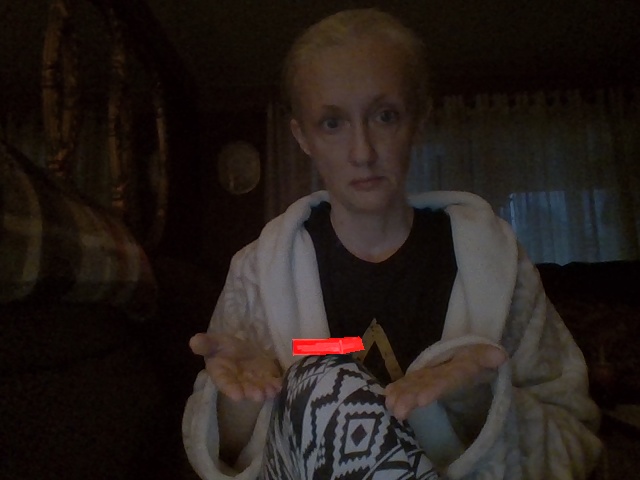}
        \caption{Ground truth.}
        \end{minipage}
    \end{subfigure}

    \begin{subfigure}[b]{\linewidth}
        \begin{minipage}{0.48\linewidth}
        \includegraphics[width=\textwidth]{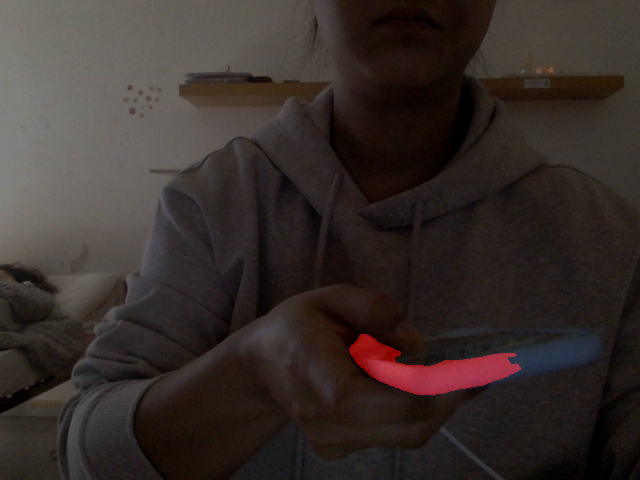}
        \caption{Prediction.}
        \end{minipage}
        \begin{minipage}{0.48\linewidth}
        \includegraphics[width=\textwidth]{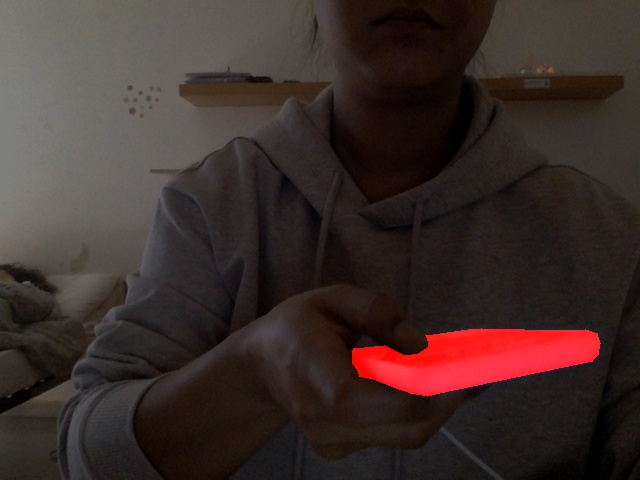}
        \caption{Ground truth.}
        \end{minipage}
    \end{subfigure}
    \begin{subfigure}[b]{\linewidth}
        \begin{minipage}{0.48\linewidth}
        \includegraphics[width=\textwidth]{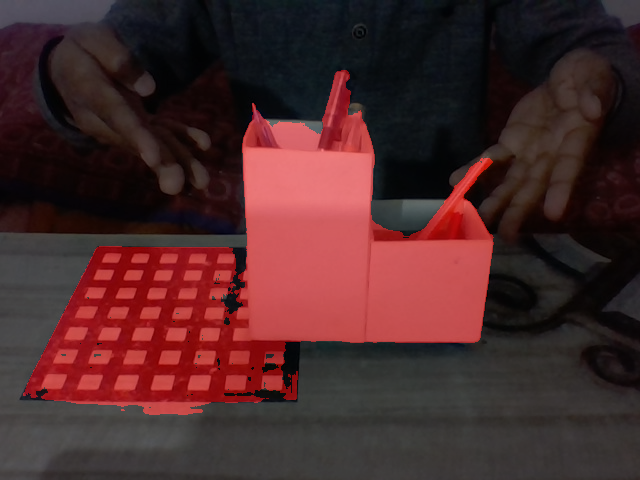}
        \caption{Prediction.}
        \end{minipage}
        \begin{minipage}{0.48\linewidth}
        \includegraphics[width=\textwidth]{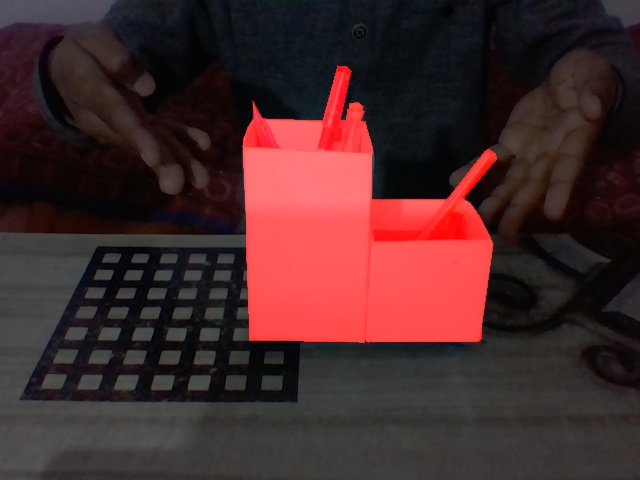}
        \caption{Ground truth.}
        \end{minipage}
    \end{subfigure}

    \caption{Additional examples of object highlight prediction failure.}
    \label{fig: fail}
\end{figure}


\subsection{Configurations of Training Object Highlights}
\label{apdx: configure}
We fine-tuned the the network of object highlights pretrained on ImageNet using the Adam optimizer for 100 epochs.
The learning rate maintains 1e-4 in the first 25 epochs, and drops exponentially for the subsequent 50 epochs until it reach 1e-5 at epoch 75.
It then maintains the learning rate of 1e-5 in the last 25 epochs.
Because there is no validation set, we simply reported the accuracy based the model achieved at the end of the training without the early stop operation.
We set the batch size to be 4.

We note that this training process  is for object highlights in \systemname.
The previous section explains how \systemname trains the model created by users (through demonstrations of objects). 
\subsection{Architecture Selections}
\label{apdx: arch_sel}

\begin{table}[t]
\caption{Performance comparison of the four models for our object highlight.}
\label{tab: arch_sel}
\begin{tabular}{ccccc}
\hline
         & \multicolumn{1}{l}{U-Net} & \multicolumn{1}{l}{UNet++} & \multicolumn{1}{l}{DeepLabV3} & \multicolumn{1}{l}{DeepLabV3+} \\ \hline
$mIoU$ & \textbf{0.718}           & 0.704                      & 0.698                         & 0.704                          \\
fps      & \textbf{28.3}            & 24.7                       & 22.5                          & 27.8                           \\ \hline
\end{tabular}
\end{table}

We chose U-Net~\cite{ronneberger2015unet}, UNet++~\cite{zhou2018unet++}, DeepLabV3~\cite{chen2017deeplab} and DeepLabV3+~\cite{Chen2018deeplabv3+} for comparison because all of them are widely used different segmentation tasks and showed good performance.
Table~\ref{tab: arch_sel} shows the results in which we compared $mIoU$ and FPS of the trained models.
The results show that U-Net was the most accurate as well as the fastest.
Note that all of the architecture used EfficientNet-b0 as the backbone.

\section{Object Highlight Prediction Failure}
\label{apdx: failure}

Figure~\ref{fig: fail} shows additional examples in which our predictions of object highlights have large discrepancy with their ground truth.

\end{document}